# Anisotropy mapping in rat brains using Intermolecular Multiple Quantum Coherence Effects

Yi Han

*Warren[2] Group, Department of Chemistry, Duke University, Durham, NC, U.S.*

E-mail: yi.han@duke.edu

3 September 2013, 5[th] version

## Abstract

This document reports an unconventional and rapidly developing approach to magnetic resonance imaging (MRI) using intermolecular multiple-quantum coherences (iMQCs). Rat brain images are acquired using iMQCs. We detect iMQCs between spins that are 10 μm to 500 μm apart. The interaction between spins is dependent on different directions. We can choose the directions on physical Z, Y and X axis by choosing correlation gradients along those directions. As an important application, iMQCs can be used for anisotropy mapping. In the rat brains, we investigate tissue microstructure.  We simulated images expected from rat brains without microstructure. We compare those with experimental results to prove that the dipolar field from the overall shape only has small contributions to the experimental iMQC signal. Because of the underlying low signal to noise ratio (SNR) in iMQCs, this anisotropy mapping method still has comparatively large potentials to grow. The ultimate goal of my project is to develop creative and effective methods of tissue microstructure anisotropy mapping. Recently we found that combining phase data of iMQCs images with phase data of modified-crazed images is very promising to construct microstructure maps. Some information and initial results are shown in this document.

# Introduction

Intermolecular Multiple Quantum Coherences (iMQCs) are unique in that they provide a fundamentally different source of anatomic and functional contrast as compared to conventional MRI. IMQCs have been shown to non-invasively probe material microstructure in liquid state NMR[1]. iMQCs contrast comes from pairs of spins separated by a well-defined and user-selectable correlation distance. Typically, only distances between 10 µm and 500 µm can be probed. This distance scale is what we refer to as mesoscopic scale. Usually, the resolution is generally limited by the available magnetic field gradient strength (spins are resolvable if the gradient separates their frequencies by more than the intrinsic linewidth), but in practice, the inherent low sensitivity and limited scan time (particularly in vivo) normally provides the more fundamental limitation. As a result, the resolution of conventional clinical MRI images is limited to voxels much larger than cellular dimensions (on the millimeter scale, typical larger than 500 µm). There is another technique called Diffusion Tensor Imaging (DTI) that probes sub-voxels effects. DTI typically probes effects smaller than 10 µm. This is because, in bulk water, molecules diffuse isotropically, with root mean square motion of approximately 7 µm in any specific direction over 10 ms. In tissues of rat brains diffusion is anisotropic, giving access to local structure on the micrometer scale (usually smaller than 10 µm). However, intermediate regimes, where the length scale ranges from around 10 µm to 500 µm are still generally difficult to access. So, being able to probe features on this mesoscopic scale makes iMQCs unique.

In this sense, iMQCs are important in that many examples of porous materials in vivo have structures on the mesoscopic scale[2]. For example, Trabecular bone consists essentially of an array of interconnected struts typically in the mesoscopic scales which form a structurally anisotropic network[3]. When bone loss which occurs in postmenopausal osteoporosis or extended exposure to microgravity happens[4], the structure and hence the degree of anisotropy and topology changes. It is associated with the progress of disease and monitoring the effects and progress of novel therapies. Conventional clinical MRI methods cannot spatially resolve the structures; DTI is also not an option because the bone pores are too large. When the material inclusions or pores are very large, it will be very time-consuming and ineffective use DTI. It is particularly inappropriate in vivo; iMQCs, on the other hand, stand out as a good method because of their ability to encode material geometry our intermediate length scales.

Besides that iMQCs signals also carry information about susceptibility differences. A fast developing imaging technique, Susceptibility Tensor Imaging[5] (STI), has been proposed and proved showing distinctive fiber pathways in 3D in the mouse brain[6]. However, the STI has a huge limitation: it requires the measurement of susceptibility at different orientations by rotating the sample with respect to the main magnetic field which is very difficult, time-consuming, and almost impossible in vivo. iMQCs, in contrast, can measure susceptibility in dependence with orientations by applying correlation gradients in different directions. This makes the experiments much easier to perform and possible in vivo.

Intermolecular multiple quantum coherence effects have many applications. One of them is anisotropy mapping. Anisotropy can be defined as a difference, in a material's magnetic properties when measured along different axes, such as conductivity, susceptibility, etc. Traditionally, anisotropy is produced by Diffusion Tensor Imaging (DTI). DTI measures the fractional anisotropy of the random motion (Brownian motion) of water molecules in the brain. Water will diffuse more rapidly in the direction aligned with the internal structure, and more slowly as it moves perpendicular to the preferred direction. This causes the anisotropy. However, iMQCs anisotropy is constructed quite differently. It is introduced in the method part below. From previous research, the iMQCs anisotropy of trabecular bone indicates the present condition of bone[2]. It is highly likely that the fractional iMQCs anisotropy in the brains can be exploited to create a map of the fiber tracts. So, anisotropy mapping is a very interesting and powerful application for iMQCs.

In this document, I will briefly introduce the dipolar field treatments which discuss the fundamental source of iMQCs for liquids or soft matter such as tissue. A set of experiments will then be discussed producing optimized intermolecular Double Quantum Coherence (iDQCs) images. I will also discuss the simulation of iDQCs signals in isotropic media. Thereafter, I explain the way to construct anisotropy maps by combining iDQCs images.. Ex vivo results in rat brains reveal that anisotropy maps in the cerebral cortex can be acquired.. Recently, we found that the anisotropy features produced by iDQCs phase data correlate much better with rat brain anatomy as compared to magnitude data. This improved correlation maybe can unveil some underlying physical effects related to the resonance frequency offset, susceptibility, etc. These effects are still under investigation. This is a critical part of my project goal. I will go to take a set of high-quality reference iMQCs images, so that the mathematical tools developed for Diffusion Tensor Imaging and Susceptibility Tensor Imaging can be used to create images that can be co-registered with those methods and anatomic images, in order to highlight what is new and different. After that, we will try to apply this technique to practical areas to obtain clinical information.

## Methods

Theoretically, for isotropic liquids it is generally assumed that the dipolar couplings between nuclear spins can be ignored because the angular dependence (1-3cos$^2$θ$_{ij}$) and the distance dependence (r$_{ij}^{-3}$) of the dipolar coupling are averaged by fast diffusion[7]. Thus, it was surprising when COSY Revamped by Asymmetric Z-gradient Echo Detection (CRAZED) was first introduced in the early 90s and showed strong iMQCs signals[8] that can only be explained when including dipolar coupled spin pairs. In order to explain the phenomenon, two basic assumptions have to be reconsidered. One is the high-temperature approximation to the Boltzmann equilibrium distribution[8,9], and the other one concerns the effect of the long range dipolar coupling[10,11,12]. Even though the correction to dipolar coupling effect in solution is still in progress[13], the principle of iMQCs is understood explicitly with these two improvements.

### Dipolar Field Treatments for Liquids

The dipolar framework starts from the Hamiltonian reflecting the dipole-dipole interaction between nuclear spins i and j

$$H_D = \frac{1}{2} \sum_{i,j} \frac{\hbar^2 \gamma_i \gamma_j}{|\vec{r}_i - \vec{r}_j|^3} \left[ \vec{I}_i \cdot \vec{I}_j - 3 \frac{(\vec{I}_i \cdot (\vec{r}_i - \vec{r}_j)) \cdot (\vec{I}_j \cdot (\vec{r}_i - \vec{r}_j))}{|\vec{r}_i - \vec{r}_j|^3} \right] \quad (1)$$

After considering the secular approximation[14] in a large external field $B_0\hat{z}$ and the different between homo-nuclear and hetero-nuclear couplings, we can write the Hamiltonian as:

$$H_D = \frac{1}{4}\sum_{i,j}\frac{\hbar^2\gamma_i\gamma_j}{|\vec{r}_i-\vec{r}_j|^3}(1-3\cos^2\theta_{rr'})\times \begin{pmatrix} [3I_{zi}I_{zj}-\vec{I}_i\cdot\vec{I}_j] & homonuclear \\ [2I_{zi}I_{zj}] & heteronuclear \end{pmatrix} \quad (2)$$

$\theta_{rr'}$ is the angle between $\vec{r}-\vec{r}'$ and $\hat{z}$. If we assume nuclear spins fast diffusion, the expectation value of $\langle 1-3\cos^2\theta_{rr'}\rangle$ should be zero over the NMR timescale (milliseconds) because each spin has sampled every $\theta_{rr'}$ around the target spin. Thus, the dipolar effects between spins separated by less than the mean diffusion length are eliminated. However, the situation is different for molecules that are much farther apart than the mean diffusion length. These spins produce a distant dipolar field (DDF):

$$\vec{B}_d(\vec{r}) = \frac{\mu_0}{4\pi}\int d^3r'\frac{1-3\cos^2\theta_{rr'}}{2|\vec{r}-\vec{r}'|^3}\left[3M_z(\vec{r}')\hat{z}-\vec{M}(\vec{r}')\right] \quad (3)$$

where now the distant dipolar field affects the time evolution as an additional field source in the Bloch equations, $d\vec{M}/dt = \gamma(\vec{M}\times\vec{B})$ where $\vec{B} = B_0\hat{z} + \vec{B}_d(\vec{r})$.

When the spins of isotropic liquid form a uniform spherical distribution, the calculated distant dipolar field from equation (3) is 0. A non-spherical sample can break the symmetry to reintroduce the distant dipolar field, even though this "shape-dependent DDF" is relatively small. In order to make the net signal measureable, this spherical symmetry has to be broken more heavily. The way we do is by creating spatially modulated magnetization, which gives a correlation distance (explained later), thus letting the sum of the dipolar interactions produce a non-zero and relatively large net effect over that distance. So applying modulation in a single direction becomes a general method to show iDQC signals. Fortunately, the dipolar field is simple and local in reciprocal space after Fourier transformation of $\vec{B}_d$ and $\vec{M}(\vec{r})$ [13]:

$$\vec{B}_d(\vec{k}) = \mu_0\left[(3(\hat{k}\cdot\hat{z})^2-1)/2\right]\left[M_z(\vec{k})\hat{z}-\vec{M}(\vec{k})/3\right] \quad (4)$$

So we choose to apply correlation distance modulation in the reciprocal space. Detailed explanation is in the next part (iDQC-Crazed Experiments). Simply saying, a gradient after 90 degree pulse can winds up the magnetization from transverse plane into a helix. The tightness of the helix, which is decided by the gradient, inherently creates a well-defined correlation distance in k space. After this, the inverse Fourier transformation gives equations below if the magnetization modulation is along a direction $\hat{s}$

$$\vec{B}_d(\vec{r}) = \mu_0 \left[(3(\hat{s}\cdot\hat{z})^2 - 1)/2\right]\left[M_z(\vec{r})\hat{z} - \vec{M}(\vec{r})/3\right] = \mu_0 \left[(3(\hat{s}\cdot\hat{z})^2 - 1)/2\right]M_z(\vec{r})\hat{z} \qquad (5)$$

In equation (5), the term $\vec{M}(\vec{r})$ can be deleted because the Bloch equation makes any terms proportional to $\vec{M}$ unaffected to the evolution. Equation (5) is for the homo-molecular situation which indicates the assumption that all of the spins have the same resonance frequency, including chemical shift. Equation (6) is for hetero-molecular situation which is between two different protons (e.g., water and acetone at normal magnetic field strengths). They don't have any transverse component in its equation because their resonance frequency differs by much more than dipolar field. The spatial integral makes it unrealistic to calculate the real DDF effect even with today's computational power[13].

$$\vec{B}_d(\vec{r}) = \frac{2\mu_0}{3}\left[(3(\hat{s}\cdot\hat{z})^2 - 1)/2\right]M_z(\vec{r})\hat{z} \qquad (6)$$

Recently, Ref 14 modified the mathematical framework of the conventional dipolar field by performing simple experiments and analyzing the corresponding unexpected results. The conclusion is that modulations in all three directions and those un-modulated components should be included in the Distant Dipole Field theory.

$$\begin{aligned}\vec{B}_d(\vec{r}) &= \frac{\mu_0}{3}\cdot\left[(2\Delta z + \Delta_\perp)M_z(\vec{r})\hat{z} + (\Delta_\perp\langle M_x\rangle\hat{x} + \Delta_\perp\langle M_y\rangle\hat{y} - 2\Delta z\langle M_z\rangle\hat{z})\right] & \text{homomolecular} \\ \vec{B}_d(\vec{r}) &= \frac{\mu_0}{3}\cdot\left[2\Delta z(M_z(\vec{r}) - \langle M_z\rangle)\hat{z}\right] & \text{heteromolecular}\end{aligned} \qquad (7)$$

For the iDQC-Crazed experiments, the two transverse magnetization components $\Delta_x$ and $\Delta_y$ have the same modulation values so in Equation (7), $\Delta z = [3(\hat{s}_z\cdot\hat{z})^2 - 1]/2$ and $\Delta_\perp = \Delta_x = \Delta_y = [3(\hat{s}_x\cdot\hat{z})^2 - 1]/2 = [3(\hat{s}_y\cdot\hat{z})^2 - 1]/2$, where $\hat{s}_x$, $\hat{s}_y$ and $\hat{s}_z$ are the direction of the modulation of initial $M_x$, $M_y$ and $M_z$, respectively. Equations (7) are the correct modifications upon Eqs. (5) and (6).

## iDQC-Crazed Experiments

The basic pulse sequence used to detect the iMQCs signals is the COSY Revamped by Asymmetric Z-gradient Echo Detection (CRAZED) pulse sequence[15 16 1 8]. In recent years, many CRAZED-like pulse sequences were designed to better show the relaxation rate and diffusion coefficients of iMQCs signals used for spectroscopic and imaging purposes. Fig.1 shows a iDQC-Crazed pulse sequence which uses a double quantum filter (GT gradient and 2GT gradient) to show intermolecular Double Quantum Coherences.

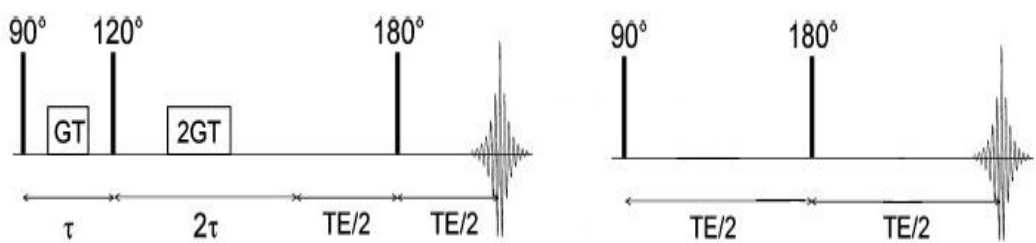

Fig. 1. The standard iDQC-CRAZED pulse sequence (left) and spin echo sequence (right)

Using a vector model analogy, the helix produced by double quantum filter is twice as tightly wound up as the single-quantum helix under a same gradient. So the process can be explained as follows: Before the first 90 degree, the quadratic term in the equilibrium density operator should be considered because it contains two-spin terms such as $I_{z1}I_{z2}$. Note that these two spin terms of the equilibrium density matrix are ignored in the high temperature approximation. A 90 degree pulse with phase –y transforms this into $I_{x1}I_{x2}$, which is actually a mixture of double- and zero-quantum coherences. Hence double-quantum coherences evolve during Tau; they also pass the double-quantum gradient filter because the second 120 degree partially transforms terms such as $I_{x1}I_{y2}$ into $I_{z1}I_{y2}$, which is a single-quantum, two-spin term. Then, $I_{z1}I_{y2}$ evolves during 2*Tau. Because the double quantum term develops at double the frequency than the single quantum term. By then, the sample exhibits periodically modulated magnetization because of the gradient pulses. Therefore, dipolar couplings between spins 1 and 2 transforms $-I_{z1}I_{y2}$ into $I_{x2}$, which is magnetization.

After using quantum point of view to explain the sequence, let's use dipolar field treatments to look at how the iDQC-Crazed signal evolves through time. We assume this is an X-crazed iDQC experiment which means we apply modulation in physical X direction. The excitation pulse puts $M_z$ into $M_x$, and then the first gradient, a part of the double quantum filter, does an X-modulation to this original longitudinal magnetization. Thereafter the 120 degree mixing pulse is applied. This mixing pulse creates a mixed term with longitudinal and transverse components on a pair of spins. Immediately, a gradient which is two times larger than the first one does an X-crazed modulation to the current transverse magnetizations. So we can see that actually the magnetization in all three directions are modulated which leads to $\langle M_\perp \rangle = \langle M_x \rangle = \langle M_y \rangle = 0$ and $\langle M_z \rangle = 0$, also based on the fact that they are both X-crazed modulated and equation (7), the $\hat{s}_\perp = \hat{x}$ and $\hat{s}_z = \hat{x}$. We can plug these values back into equation (7) and get the evolution results below:

$$\begin{aligned}\vec{B}_d(\vec{r}) &= \mu_0 \cdot \Delta \cdot M_z(\vec{r})\hat{z} & \textit{homomolecular} \\ \vec{B}_d(\vec{r}) &= \frac{2\mu_0}{3} \cdot \Delta \cdot M_z(\vec{r})\hat{z} & \textit{heteromolecular}\end{aligned} \quad (9)$$

For the iDQC-Crazed sequence, the homo-molecular signal and the hetero-molecular signal have similar forms which are shown in equation (9), so we don't have to consider the simulation difference between these two signals.

In our experiments, we scan our sample, rat brains, at 7T. They are doped with gadolinium. The correlation distance (related to the gradient pair) here we applied is 70 m (explained later). The 90 degree and 120 degree pulses are Gaussian pulses. Adiabatic hyperbolic secant pulsed were used for two refocusing 180 degree pulses. The repetition time for all images was 1 s which is much larger than T1 of rat brains which was measured to be 40ms. The slice thickness was 2 mm with a 3cm*3cm FOV. The acquisition matrix size was 256*256.

Before taking iDQC scans, spin echo images were also acquired. The right picture in Fig. 1 shows the sequence. The 90 degree and 180 degree pulses were Gaussian pulses. The repetition time for all images was 14 s. The echo time was 10 ms for rat brains. The slice thickness was 2 mm with a 3cm*3cm field of view. The spin echo acquisition matrix size is 256*256.

## iDQC-Crazed Simulation

The signal in a typical Crazed sequence[17][10] is approximately equal to the normal magnetization (proportional to proton density) multiplied by the dipolar field. This formula can be explained from:

$$\frac{d\vec{M}(\vec{r},t)}{dt}=\gamma\vec{M}(\vec{r},t)\times[B_0\hat{z}+\vec{B}_d(\vec{r},t)] \tag{10}$$

Specifically, $\vec{M}(\vec{r},t)$ on the right of equation (10) can be calculated by applying a modulation to the magnetization signal in the frequency domain. The modulation here means a certain correlation distance ($d_c$) in frequency domain determined by:

$$d_c=\frac{\pi}{\gamma GT} \tag{11}$$

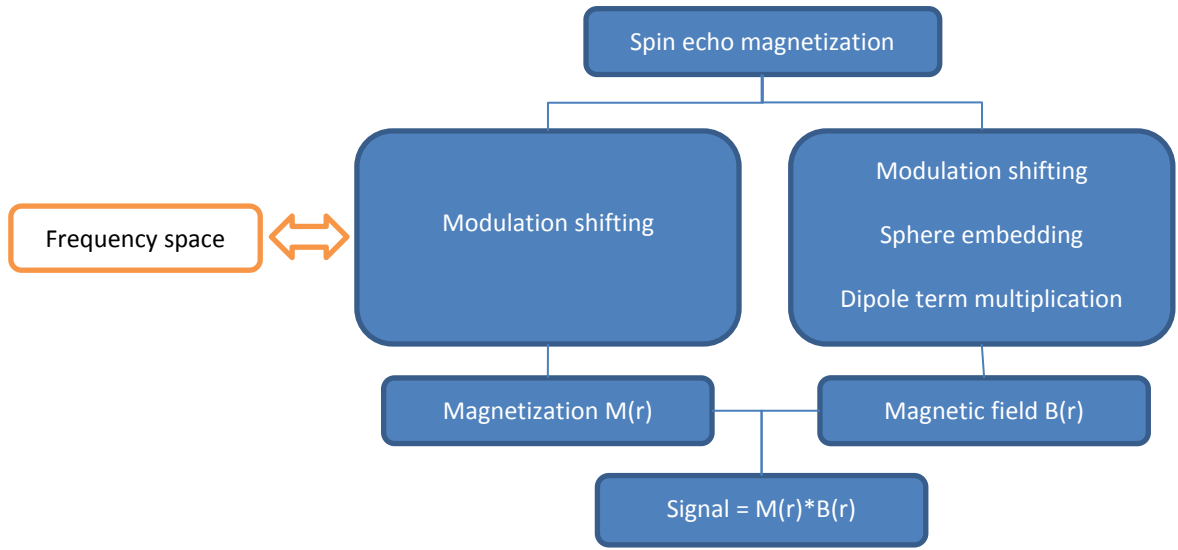

Fig. 2. The process of iDQC simulation: Input the spin echo magnetization, perform modulation in the k space including uniform dipole term multiplication, Fourier transform back those images and construct simulated iMQCs.

$B_d(\vec{r})$, which is the dipolar field, depends on the magnetization distribution and thus on the shape of the object[18][13]. The uniform dipolar field can be found by Fourier transformation of the magnetization density, multiplying by the uniform dipolar term $(3(\hat{k}\cdot\hat{z})^2-1)/2$ [19], followed by an inverse Fourier transform. This can easily be performed for any shape given a high resolution image that provides the magnetization distribution. However, before multiplying the dipolar term, we should apply the same modulation to $\vec{M}(\vec{r},t)$, and then embedding a sphere. The step "Sphere embedding" is not only for standardization of the frequency domain images for uniform comparisons between images, but also for transferring the information of the center point to other points without causing any distortions to the dipolar field. The center point of dipole term in frequency domain is a singular point. Its value becomes 0 by embedding a sphere symmetrically, at the same time, the singularity goes to infinite location.

**Anisotropy mapping with iDQCs**

Understanding iMQCs' origin and having its images, it is now becoming more and more interesting to use these to construct anisotropy images. Previous work has shown that multiple quantum anisotropy imaging, performed by iMQCs images with correlated gradients in three perpendicular directions, can detect differences in structured materials (such as tumors with embedded nanoparticles)[15][4][2].

For typical imaging applications, iMQCs images are more sensitive to the susceptibility interfaces than spin echo images. By varying the direction and amplitude of the correlation gradient pairs, unique contrast can be obtained. The direction of the correlation gradient pair affects the contrast because the dipolar field causes a spatial dependence of the iMQCs signal that is proportional to $(3\cos^2\theta\text{-}1)/2$, where $\theta$ is the angle between $B_0$ and the direction of the correlation gradient. As a result, the signal for a Z correlation gradient should be the opposite sign and two times larger than the signals for X and Y correlation gradients. In isotropic media, adding the three complex images together or subtracting the magnitude images should yield 0. On the other hand, in anisotropic media, these combinations reflect the local structure. The strength of the correlation gradients determines the distance between the two coupled spins. In isotropic media, the correlation distance should have no effect on the signal (ignoring the diffusion and sample size limitations). For anisotropic media, however, the correlation distance affects the image contrast because the signal is directly related to the dipolar field at that distance.

We knew that the iDQC signal formula contains many terms[17], such as contributions from dipolar field, T2, the resonance frequency offset, the dipolar demagnetizing time ($\tau_d = (\gamma\mu_0 M_0)^{-1}$), pulse flip angles, etc. But roughly, |Z++Y+X| and |Z|-|Y|-|X| should be 0 for isotropic areas and show signals for anisotropic areas. For the simulation, generally only uniform dipolar field is considered which means that |Z++Y+X| and |Z|-|Y|-|X| should be very close to 0.

Previous studies have shown that mesoscopic structural anisotropy maps can be obtained with iMQCs[4, 2, 20, 21].

## Result/Discussion

Before directly showing the anisotropy maps, we would like at first to compare the intermolecular multiple quantum coherences images with different kinds of other conventional MRI images (Fig.3). Clearly, the contrast rising from iDQC is more obvious than any other conventional MRI shown here. More important, unlike conventional MRI, where image contrast is largely based on variations in spin density and relaxation times (often with injected contrast agents), contrast with iDQC images comes from dipolar couplings in intermediate scales dictated by gradient strength. In the rapidly expanding field of functional MRI, contrast is frequently the limiting factor. New methods for contrast enhancement could thus improve tissue characterization, particularly if they correlate with physiologically important characteristics. It has already been reported that this contrast is useful in the detection of small tumors[20] and in functional MRI[2, 4].

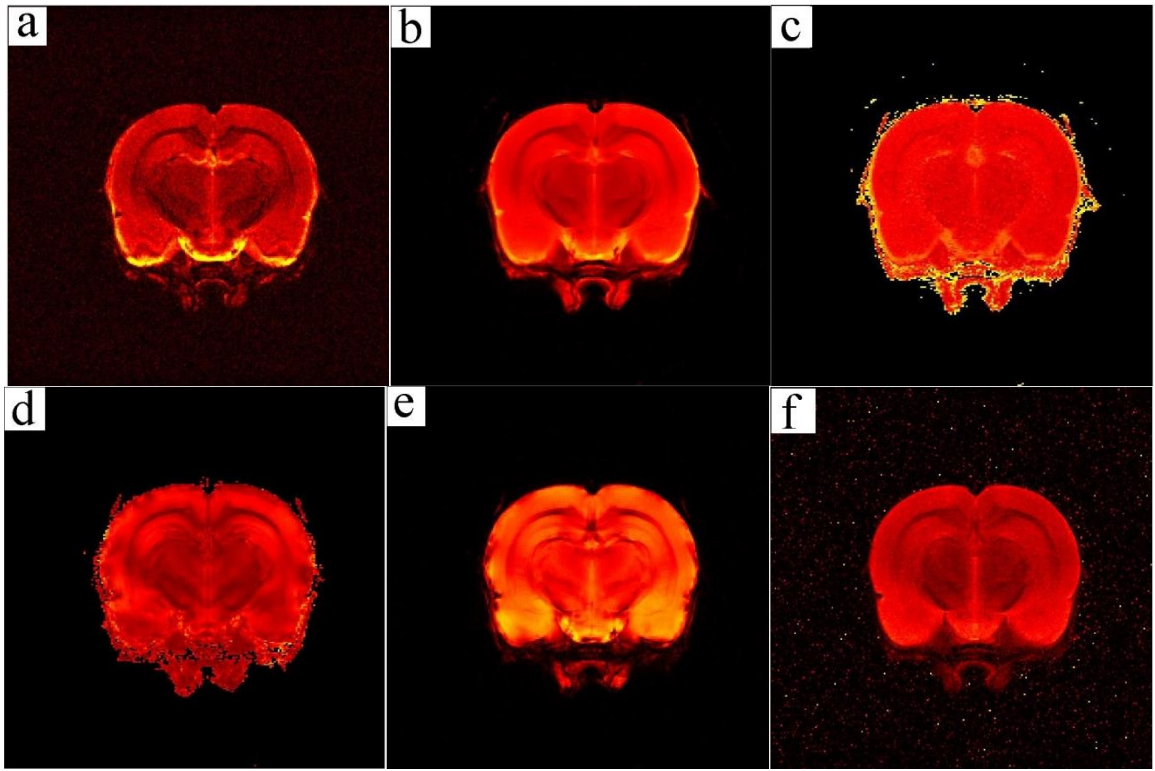

Fig. 3. All the images are rat brain images scanned at 7T, the scanning thickness is 2mm, the field of view is 2.5cm*2.5cm. (a). iDQC-Crazed image (b). Spin echo image (c). T2 map (d). T2* map (e). Proton density map (f). Diffusion trace weighted image (It is calculated as s0*exp(-b*trace), s0 is the diffusion experiment signal without gradient, b is a constant related to diffusion gradient, The trace is calculated after diagonalization to the diffusion tensor)

Then we can go to look at this very important application which is anisotropy mapping. Any MRI sequence contains 3 channels which are slice selection channel, phase encoding channel and frequency encoding channel. For the iDQC sequence introduced above, we can put the double quantum filter (gradients GT, 120 degree pulse and 2GT) in different channels which are, the slice selection channel (physical Y direction), the phase encoding channel (physical X direction) and the frequency encoding channel (physical Z direction). Thus, iDQC images in different physical directions are obtained (top images of Fig. 4). From the general intensity and color scales, it is easy to confirm that the intensity of the Z-crazed image is roughly twice that of Y-crazed or X-crazed, which proves that the experiments accords with the dipolar field term $(3\cos^2\theta-1)/2$. The bottom images of Fig (4) show the anisotropy. The |Z+Y+X|/max (|Z|) map contains phase information while (|Z|-|Y|-|X|) /max (|Z|) doesn't. The anisotropy map

without phase information seems to have larger contrast. But anyway, the anisotropy maps show some features which can be seen very easily. (Such as the bottom part which has a very large signal) Right now, we can't extract any clinical information from those anisotropy maps. To relate iDQC images and anisotropy maps with clinical conditions definitely can be one of my future work.

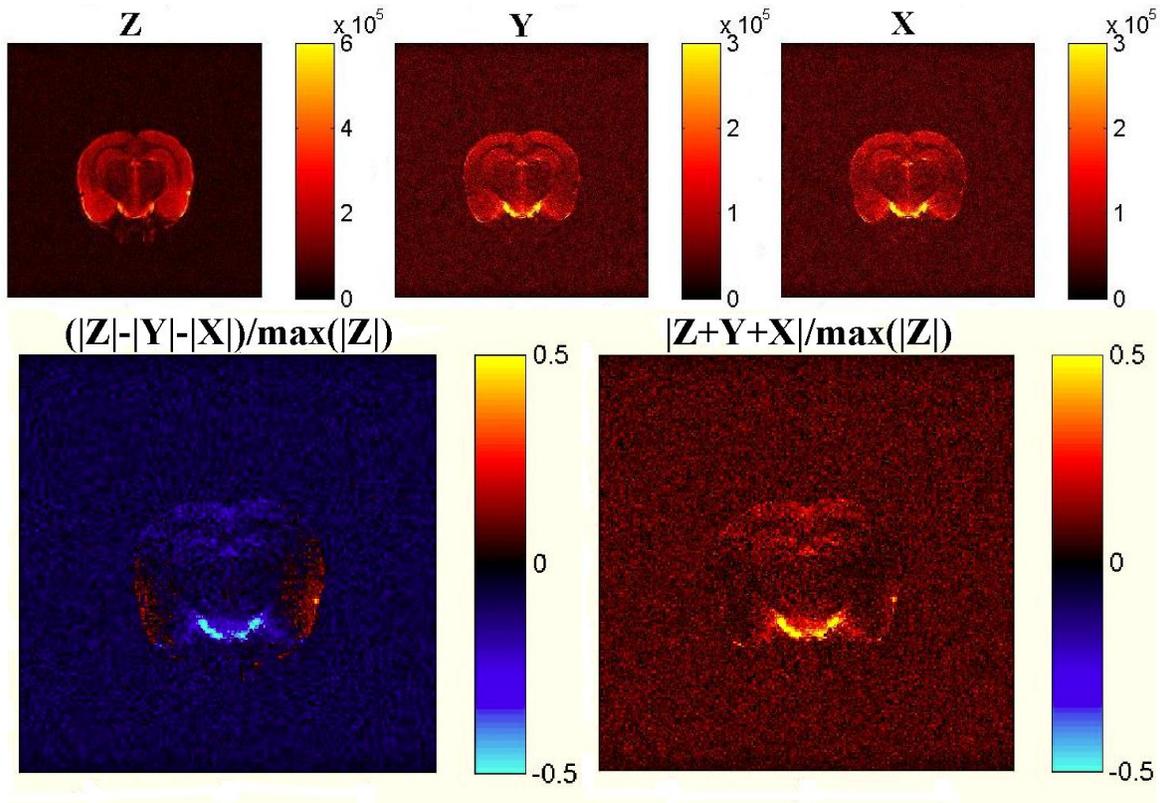

Fig. 4. iDQC-Crazed rat brain experimental images. The main magnetic field direction is along Z direction. The top row is the intensity maps produced by considering correlation distance in 3 directions Z, Y and X. The bottom row displays the fractional anisotropy maps which are calculated by (|Z|-|Y|-|X|)/max (|Z|) and |Z+Y+X|/max (|Z|).

Fig (5) shows the simulated fractional magnitude anisotropy images, (|Z|-|Y|-|X|)/max (|Z|) and |Z+Y+X|/max (|Z|). These simulations prove that the theory of iDQC images and corresponding anisotropy measurements are correct and could be very powerful. The intensity of Z-Crazed is roughly twice of that of Y-Crazed and X-Crazed which proves that the dipolar field term $(3\cos^2\theta-1)/2$ is working here. More important result we want to see is that when you do (|Z|-|Y|-|X|)/max (|Z|) and |Z+Y+X|/max (|Z|) using simulated data for the anisotropy maps,

there should be no signals at all because the uniform dipole field is the only effect here. This is different in experiments, in iDQC experiments the dipole field term plays a leading role but still there will be some residual signals which are located in anisotropic areas. Anyway, when we go back to simulation, theoretically speaking we should see no anisotropy for uniform dipole field situation, but here we still see tissue microstructures in the rat brain in the bottom images of Fig (5). Is it wrong? Actually, this is just a trick of choosing different color-bar. In Fig (6), I choose the same color-bar for experimental and simulated images which is easier to compare. Clearly the conclusion is that no fractional anisotropy is detected for uniform dipolar field, which means the simulation we did is correct.

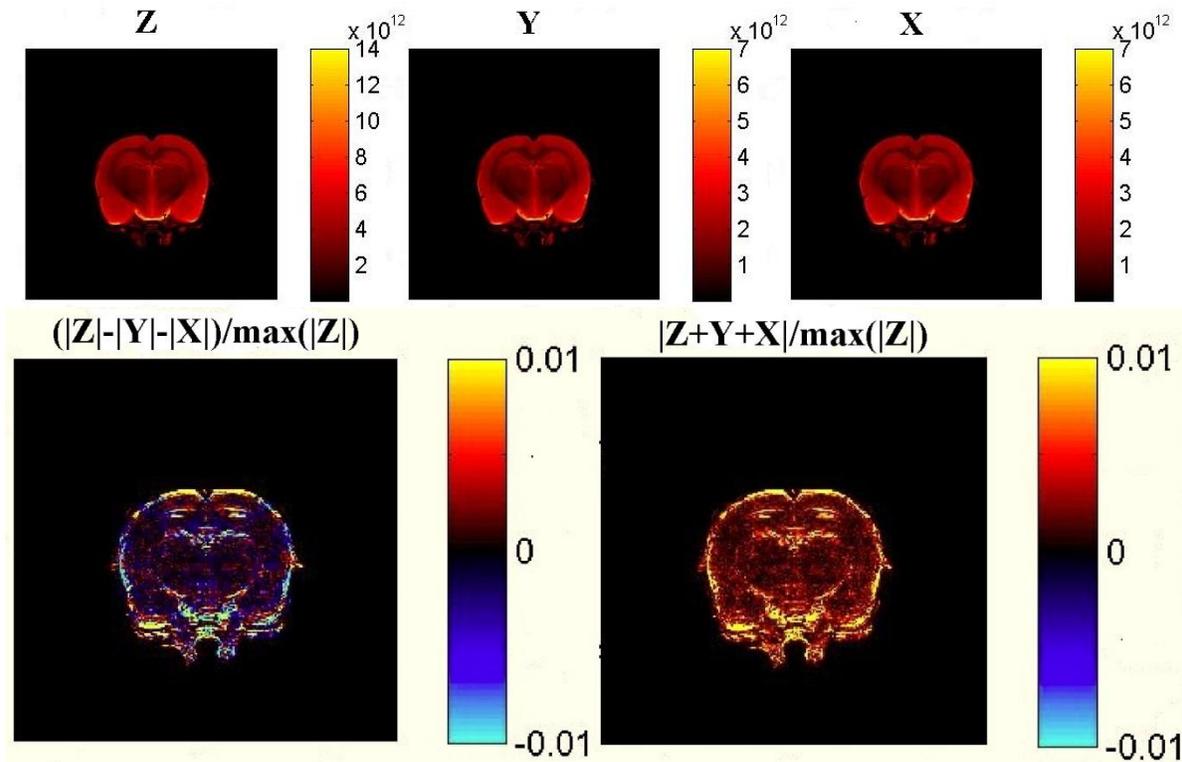

Fig. 5. iDQC-Crazed rat brain simulated images. The top row displays the simulated images which are produced applying correlation distance to spin echo density (S) weighted map in 3 directions, Z, Y and X. The bottom row displays the fractional anisotropy maps (|Z|-|Y|-|X|)/max (|Z|) and |Z+Y+X|/max (|Z|) a by uniform dipole field term.

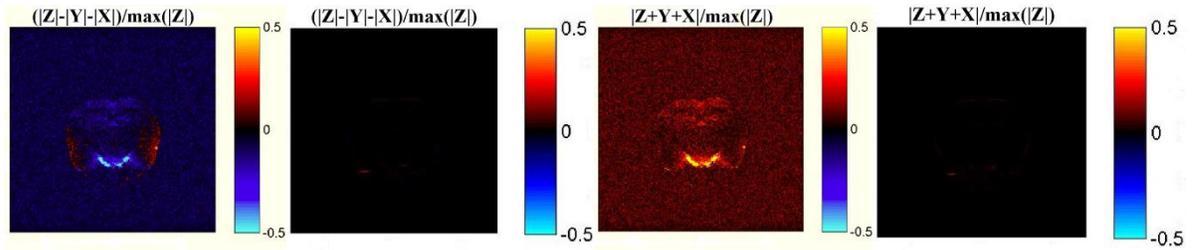

Fig 6. Experimental fractional anisotropy maps and simulated fractional anisotropy maps in the same color-bar. The first and third images are experimental (|Z|-|Y|-|X|)/max (|Z|) map and |Z+Y+X|/max (|Z|) map. The second and fourth images are simulated (|Z|-|Y|-|X|)/max (|Z|) map and |Z+Y+X|/max (|Z|) map.

# Future work

## iDQC effects on different correlation distance

The first future work is to look at the effect from different correlation distance to iDQC images and corresponding anisotropy maps. The correlation distance used in above images is 70 μm. It is possible that another correlation distances will better enhance the signal contrast and better show anisotropy information in the rat brain. The correlation distance can be changed by the correlation gradients strength and duration based on equation (11). Besides rat brains, it is highly possible that some other biological information or diseases only can be better achieved or studied by using a very narrow correlation distance. So by applying different correlation distance in mesoscopic scales, the produced iDQC images should be studied qualitatively and quantitatively.

## iDQC fractional phase anisotropy and susceptibility

Besides using magnitude data to construct anisotropy images, we are still trying to find other ways or improvements to construct anisotropy and explain its physics basis. After several new trials of experiments, we found that fractional phase anisotropy maps is every clear and informative. They are produced by following steps: take the phase data of the signals and choose a small area of rat brain which is very flat and isotropic, then match the values of that area in Z, Y, X to be 2:-1:-1 by multiplying constants, and at last apply those constants to the whole images not just that small and isotropic area. At the first glance this method seems like having no physics meaning. However, considering that the iDQC phase data actually has physical basis which is related to susceptibility (explained later), it is likely that the phase data intensity in dependence with correlation distance orientations can be clearly studied. More importantly, in Fig (5), most areas of rat brain are similar to the background, but some anisotropic areas show signals. So, these images suggest that the phase information which is unimportant in the traditional treatment can actually clearly give new contrast. For example, we can color-code the phase FA map and compare the contrast with conventional Diffusion Tensor Imaging and other MRI images. In any case, there is plenty of new information to co-register with other methods, and the importance is that it operates on a distance scale (tens of microns) which is more relevant for understanding tissue morphology.

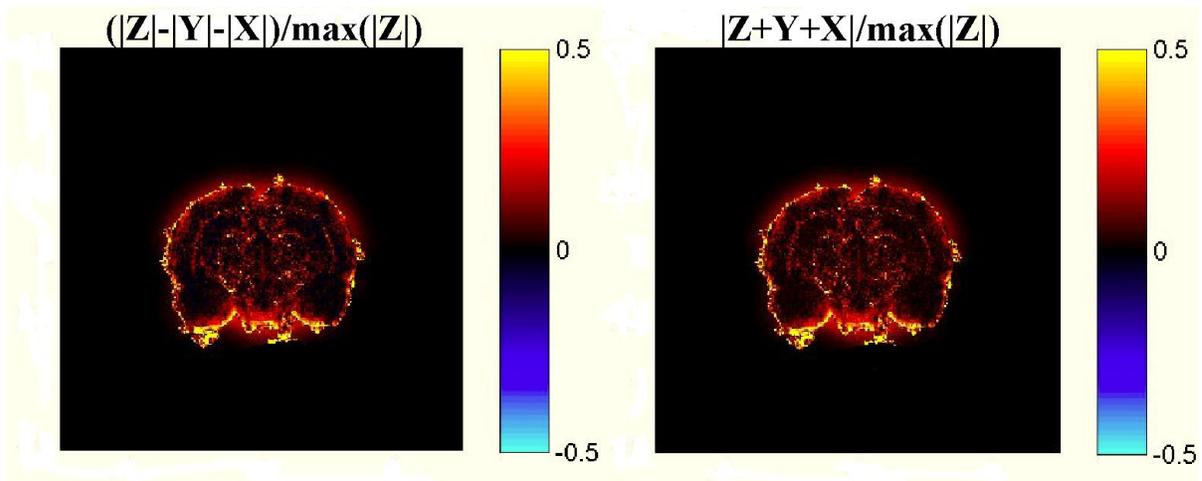

Fig. 7. Experimental fractional phase anisotropy maps, (|Z|-|Y|-|X|)/max (|Z|) and |Z+Y+X|/max (|Z|)

Another perspective to look at the phase information is to think about what we can get by changing the standard iDQC sequence. Fig (8) borrows the first row images from Fig (3) which are produced by Standard-Crazed sequence. They have contrast from both magnetization density and resonance frequency variation; so-called modified-CRAZED sequences, with an extra 180 degree pulse in the middle of Tau interval, can produce images which only have contrast from magnetization density variation. So the difference between Standard-Crazed images and Modified-Crazed images from the XY terms of the dipolar field predicts local phase shifts from susceptibility or magnetization anisotropy. This has prompted a more careful examination, borrowing the tools recently developed for phase interpretation in susceptibility weighted imaging and susceptibility tensor imaging. The raw phase from such images is dominated by unwanted artifacts such as coil phase shifts and shim effects. Thus the observed phase requires several levels of correction (phase unwrapping, fitting and deconvolution) and projection onto a dipole field which assumes susceptibility proportional to magnetization.

It also drops a hint that this perspective is very promising when comparing the T2 map and T2* map in fig (3). Clearly, their contrast is very different which means that the magnetic field inhomogeneities and susceptibility effects play an essential role in T2* map. So digging into the susceptibility-related images makes perfect sense.

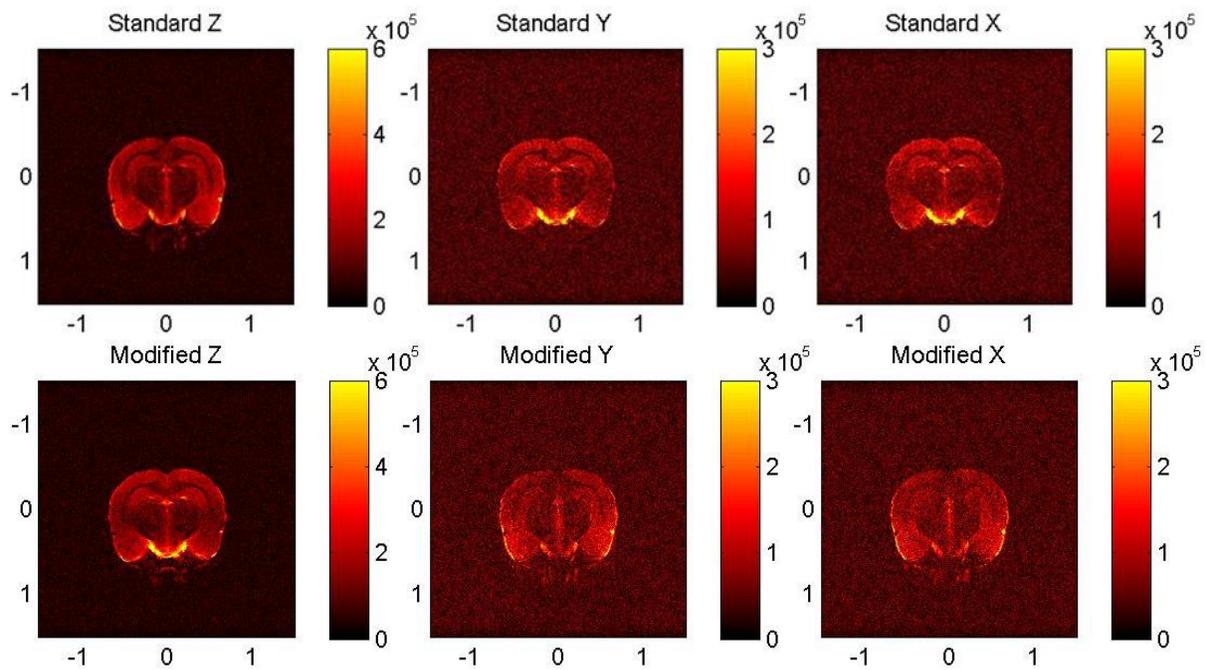

Fig. 8. Standard-Crazed and Modified-Crazed rat brain images. The top row is the intensity maps produced by applying correlation distance in 3 directions, Z, Y and X. The bottom row is the corresponding modified maps which insert 180 degree pulse in the middle of Tau interval. You can clearly see the different contrasts between these two sets of images. Notice that standard crazed sequences and modified crazed sequences have the same timing after 120 mixing pulse.

To sum up the second future work above, it should be to better understand fractional phase anisotropy maps and the difference between Standard-Crazed and mod-Crazed images. Clearly, there are still many artifacts in those images which can be improved, such as coil phase shifts and shim effects. We will work with Chunlei Liu at CIVM who has recently shown that diffusion and susceptibility MRI provide complementary information of white matter microstructure. More importantly, they are new methods and these phase-relevant results show a total new contrast which can be co-registered with many other MR methods, and there are many applications in the future because they are looking at the proper scales 10 μm to 500 μm.

### iDQC fractional phase anisotropy and susceptibility: practical examples

Definitely another important part of my future work is to apply the intermolecular multiple quantum coherences effect into clinical research when the theory of my second future work is built, which is about phase data information and susceptibility. Considering that it is already said in the introduction part that iDQC images have huge advantages over susceptibility weight imaging technique, we should really use iDQC effects to do susceptibility-related practical research. So, what susceptibility-related images can measure, iDQC images can measure that, too. For examples, brain iron concentrations in vivo[22] and unprecedented anatomical contrast in both white and gray matter regions[6, 23]. The clinical potential of susceptibility-related images is still under investigation but it is anticipated that it will provide novel insights into disease induced tissue change[24]. Thus, it is highly possible that intermolecular multiple quantum coherences images can also give us those insights without suffering the difficulty of susceptibility tensor imaging experiments.